# Simplified vibratory characterization of alluvial basins

### Jean-François SEMBLAT [a], Roberto Paolucci [b], Anne-Marie DUVAL [c]

[a] Laboratoire Central des Ponts et Chaussées, 58 bd Lefebvre, 75732 Paris Cedex 15, France, semblat@lcpc.fr
[b] Politecnico di Milano, P.za Leonardo da Vinci 32, 20133 Milano, Italy
[c] CETE Méditerranée, Laboratoire de Nice, ERA Risque sismique, 56 bd de Stalingrad, 06300 Nice, France

**Abstract.** *For the analysis of seismic wave amplification, modal methods are interesting tools to study the modal properties of geological structures. Modal approaches mainly lead to information on such parameters as fundamental frequencies and eigenmodes of alluvial basins. For a specific alluvial deposit in Nice (France), a simplified modal approach involving the Rayleigh method is considered. This approach assumes a set of admissible shape functions for the eigenmodes and allows a fast estimation of the fundamental frequency of the basin. The agreement between modal numerical results and experimental ones is satisfactory. The simplified modal method then appears as an efficient mean for the global vibratory characterization of geological structures towards resonance.*

*Caractérisation vibratoire simplifiée de remplissages sédimentaires*

**Résumé.** *L'analyse de l'amplification du mouvement sismique peut être réalisée en étudiant les propriétés vibratoires des structures géologiques. Les approches modales peuvent donner des informations telles que les fréquences propres et les modes propres de bassins sédimentaires. Pour étudier la résonance d'un bassin sédimentaire du centre de Nice, une approche modale simplifiée utilisant la méthode de Rayleigh est proposée. Cette approche suppose la connaissance d'une famille de fonctions modales admissibles et permet une détermination rapide de la fréquence fondamentale du bassin. La comparaison avec d'autres résultats numériques (obtenus par équations intégrales de frontière) et des mesures réalisées sur site est très favorable. Cette méthode conduit donc aisément et rapidement à une caractérisation vibratoire globale de structures géologiques vis-à-vis de la résonance.*

*Version française abrégée*

L'amplification locale du mouvement sismique dans les bassins sédimentaires peut parfois être très importante [1,2,7,10,11]. La propagation des ondes sismiques dans les structures géologiques de surface peut en effet conduire à une forte amplification du mouvement due au contraste de propriétés mécaniques entre un bassin sédimentaire et le substratum rocheux. L'amplification du mouvement sismique peut également s'interpréter comme la résonance des couches de surface à certaines fréquences. Les effets de site peuvent alors être analysés en termes de résonance vibratoire des structures géologiques de surface [8,12]. Dans cette note, nous considérons une approche modale simplifiée basée sur la méthode de Rayleigh [9] pour estimer directement la fréquence fondamentale d'une structure géologique.

La méthode modale considérée ici vise à estimer la fréquence « fondamentale » d'une structure géologique. Le processus de propagation d'ondes conduit généralement à une forte amplification du mouvement à des fréquences variées [11] mais, dans un but pratique, il est intéressant de déterminer une fréquence fondamentale unique par des approches simplifiées. La méthode de Rayleigh peut en effet permettre une estimation rapide et fiable de la fréquence de résonance d'une structure géologique [9].

Cette méthode considère que le déplacement d'un système élastique dans un de ses modes propres peut être approché par celui d'un système à un degré de liberté. Nous étudions ici le premier mode propre caractérisé par la fréquence : $\omega_0=2\pi f_0$. En notant $V$ l'énergie élastique du système et $T$ son énergie cinétique, la conservation de l'énergie totale d'un système élastique implique que $V_{max}=T_{max}$. On peut alors écrire le déplacement $s_k(\underline{x},t)$ correspondant aux vibrations harmoniques à la fréquence $\omega_0$. La valeur exacte de la fréquence fondamentale du système peut ainsi être obtenue lorsque la forme modale réelle $\psi_k(\underline{x})$ est connue. Cependant, comme la solution exacte n'est généralement pas accessible, la valeur de $\omega_0$ peut être estimée correctement en considérant une approximation réaliste de $\psi_k(\underline{x})$. Cette approximation doit satisfaire à la fois les compatibilités géométriques et les conditions aux limites. Il a néanmoins été démontré [9] que la seconde condition peut ne pas être satisfaite complètement et que la forme modale peut être choisie dans une large gamme de fonctions satisfaisant uniquement les compatibilités géométriques. On peut alors déterminer la fréquence fondamentale à l'aide de la relation suivante :

$$\omega_o^2 \leq \min_{\psi_k} \frac{\int_\Omega \sigma_{jl}(\underline{x})\varepsilon_{jl}(\underline{x})d\Omega}{\int_\Omega \rho(\underline{x})\psi_x^2(\underline{x})d\Omega} \qquad (1)$$

où $\sigma_{ij}$ est le tenseur de contrainte, $\varepsilon_{ij}$ le tenseur de déformation, $\rho$ la masse volumique et $\psi_k(\underline{x})$ la forme modale.

En vue de l'analyse vibratoire simplifiée, nous considérons un bassin sédimentaire situé dans le centre de Nice. Différents types de mesures (bruit de fond, mouvements sismiques faibles) ont permis d'analyser l'amplification des ondes sismiques dans le bassin principal [7]. Pour évaluer la robustesse de la méthode, le bassin est supposé homogène et sa résonance vibratoire est analysée pour une sollicitation sismique antiplane en cisaillement caractérisée par la célérité $V_S = \sqrt{\mu/\rho}$. Les caractéristiques mécaniques des deux milieux (bassin et substratum) sont les suivantes :
- bassin : $\rho_1$=2000 kg/m$^3$, $\mu_1$=180 MPa, $\nu$=0.2 c'est-à-dire $V_S$=300 m/s ;
- substratum : $\rho_1$=2300 kg/m$^3$, $\mu_1$=4500 MPa, $\nu$=0.2 d'où $V_S$=1400 m/s.

où $\rho$ est la masse volumique, $\mu$ le module de cisaillement, $\nu$ le coefficient de Poisson et $V_S$ la célérité des ondes de cisaillement.

Comme le montre la figure 1, l'interface entre le bassin et le substratum est décrite à l'aide de deux fonctions cosinus (parties ouest et est de la vallée). A partir de formes modales admissibles choisies judicieusement, on estime les fréquences fondamentales de ces deux parties du bassin sédimentaire. Les résultats sont obtenus pour le premier mode en considérant différentes valeurs du module de cisaillement : $\mu_1$=180 MPa (i.e. $V_S$=300 m/s), $\mu_2$=120 MPa (i.e. $V_S$=245 m/s) et $\mu_3$=90 MPa (i.e. $V_S$=212 m/s). Les résultats numériques obtenus par la méthode de Rayleigh sont comparés aux fréquences d'amplification maximale obtenues expérimentalement et numériquement (méthode des éléments de frontière).

D'après ces résultats, nous pouvons constater que les fréquences fondamentales obtenues par la méthode de Rayleigh sont en bon accord avec les fréquences de forte amplification obtenues expérimentalement et par approche propagative. Les valeurs données dans le tableau I sont comparées pour la résonance de la partie profonde du bassin (ouest) et la partie peu profonde (est) et ce pour différentes valeurs du module de cisaillement. Pour un modèle de bassin homogène, la méthode de Rayleigh semble donc donner des résultats très intéressants qui sont non seulement en accord avec les valeurs obtenues à l'aide d'autres modèles mais également avec les résultats expérimentaux (figure 2).

## 1. Introduction

The local amplification of seismic motion in alluvial basins could sometimes be large [1,2,7,10,11]. Seismic wave propagation in surface geological structures can indeed lead to a strong motion amplification due to the impedance constrast between the alluvial deposit and the bedrock. However, the amplification of seismic motion can also be considered as a result of the resonance of surface layers at peculiar frequencies. Site effects can then be analysed in terms of vibratory resonance of surface geological structures [8,12]. In this note, we consider a simplified modal approach based on the Rayleigh's method [9] for the direct estimation of the fundamental frequency of a geological structure.

The modal method considered herein aims at estimating the « fundamental » frequency of an alluvial basin. The wave propagation process generally leads to strong motion amplifications at various frequencies [11] but, from a practical point of view, it is interesting to estimate a unique fundamental frequency by a simplified efficient method. We will show herein that Rayleigh's method allows a fast and reliable estimation of the resonance frequency of a geological structure [9].

## 2. Simplified modal method

### 2.1 Theoretical point of view

The modal method considered herein vise à estimer la fréquence « fondamentale » d'une structure géologique. Le processus de propagation d'ondes conduit généralement à une forte amplification du mouvement à des fréquences variées (Semblat et al 00b) mais, dans un but pratique, il est intéressant de déterminer une fréquence fondamentale unique par des approches simplifiées. La méthode de Rayleigh permet par exemple une estimation rapide et fiable de la fréquence de résonance d'une structure géologique (Paolucci 99).

Cette méthode considère que le déplacement d'un système élastique dans un de ses modes propres peut être approché par celui d'un système à un degré de liberté. Nous étudions ici le premier mode propre caractérisé par la fréquence : $\omega_0 = 2\pi f_0$. En notant $V$ l'énergie élastique du système et $T$ son énergie cinétique, la conservation de l'énergie totale d'un système élastique implique que $V_{max} = T_{max}$. Le déplacement $s_k(\underline{x}, t)$ correspondant aux vibrations harmoniques à la fréquence $\omega_0$ peut s'écrire :

$$s_k(\underline{x}, t) = \psi_k(\underline{x}) e^{i\omega_o t} \qquad (2)$$

où $\underline{x}$ représente la coordonnée d'espace, $i$ le nombre imaginaire, $t$ le temps et $\psi_k(\underline{x})$ la forme modale suivant la direction $k$. On calcule alors l'énergie cinétique du système comme suit :

$$T(t) = \int_\Omega \frac{1}{2} \rho(\underline{x}) \left( \frac{\partial s_k}{\partial t} \right)^2 d\Omega = -\omega_0^2 e^{2i\omega_o t} \int_\Omega \frac{1}{2} \rho(\underline{x}) \psi_k^2(\underline{x}) d\Omega \qquad (3)$$

d'où

$$T_{max} = \max_t T(t) = -\omega_o^2 \int_\Omega \frac{1}{2} \rho(\underline{x}) \psi_k^2(\underline{x}) d\Omega \qquad (4)$$

L'énergie élastique $V$ s'écrit alors :

$$V(t) = \int_\Omega \frac{1}{2} \sigma_{jl}(\underline{x}) \varepsilon_{jl}(\underline{x}) d\Omega \qquad (5)$$

où $\varepsilon_{ij} = s_{i,j} + s_{j,i}$ est le tenseur de déformation et $\sigma_{ij} = \lambda \varepsilon_{ll} \delta_{ij} + 2\mu \varepsilon_{ij}$ le tenseur de contrainte obtenu par la loi de Hooke avec $\delta_{ij}$ le symbole de Kronecker, $\lambda$ et $\mu$ les coefficients de Lamé. Comme pour $T_{max}$, $V$ atteint sa valeur maximale quand $|e^{2i\omega_o t}| = 1$. Il vient ainsi :

$$\omega_o^2 = \frac{\int_\Omega \sigma_{jl}(\underline{x})\varepsilon_{jl}(\underline{x})d\Omega}{\int_\Omega \rho(\underline{x})\psi_k^2(\underline{x})d\Omega} \qquad (6)$$

L'équation 6 donne la valeur exacte de la fréquence fondamentale du système lorsque la forme modale réelle $\psi_k(\underline{x})$ est connue. Cependant, comme la solution exacte n'est généralement pas accessible, la valeur de $\omega_0$ peut être estimée correctement en considérant une approximation réaliste de $\psi_k(\underline{x})$. Cette approximation doit satisfaire à la fois les compatibilités géométriques et les conditions aux limites. Il a néanmoins été démontré (Paolucci 99) que la seconde condition peut ne pas être satisfaite complètement et que la forme modale peut être choisie dans une large gamme de fonctions satisfaisant uniquement les compatibilités géométriques. On peut alors déterminer la fréquence fondamentale à l'aide de la relation suivante :

$$\omega_o^2 \leq \min_{\psi_k} \frac{\int_\Omega \sigma_{jl}(\underline{x})\varepsilon_{jl}(\underline{x})d\Omega}{\int_\Omega \rho(\underline{x})\psi_x^2(\underline{x})d\Omega} \qquad (7)$$

## 2.2 Estimation of the fundamental frequency of a geological structure

En vue de l'analyse vibratoire simplifiée, nous considérons un bassin sédimentaire situé dans le centre de Nice. Ce site est bien connu puisque de nombreuses expérimentations ont été réalisées ainsi que plusieurs modélisations numériques (Semblat et al 00b). Dans ce texte, le bassin est supposé homogène et sa résonance vibratoire est analysée pour une sollicitation sismique antiplane en cisaillement caractérisée par la célérité $V_S = \sqrt{\mu/\rho}$. Les caractéristiques mécaniques des deux milieux (bassin et substratum) sont les suivantes :
- bassin : $\rho_1$=2000 kg/m$^3$, $\mu_1$=180 MPa, $\nu$=0.2 c'est-à-dire $V_S$=300 m/s ;
- substratum : $\rho_1$=2300 kg/m$^3$, $\mu_1$=4500 MPa, $\nu$=0.2 d'où $V_S$=1400 m/s.

où $\rho$ est la masse volumique, $\mu$ le module de cisaillement, $\nu$ le coefficient de Poisson et $V_S$ la célérité des ondes de cisaillement.

Comme le montre la figure 1, l'interface entre le bassin et le substratum est décrite à l'aide de deux fonctions cosinus :
- partie ouest de la vallée : $f(x,z) = (h_1+1).\cos(2,7.10^{-3} x+1,55)$
- partie est de la vallée : $g(x,z) = (h_2+2).\cos(2,8.10^{-3} x -1,3)$

Parmi les formes modales admissibles, on choisit les suivantes (Paolucci 99) :

$$\psi_2(x,z) = \cos^r\left(\frac{\pi}{2}(1-f(x,z))\right) \times \sin^{2s+1}\left(\frac{(n+1)\pi}{2}\left(1+\frac{x}{a}\right)\right) \times \cos^t\left(\frac{(2m+1)\pi}{2}\frac{z}{h}\right) \qquad (8)$$

où $f(x,z)$ est la fonction donnée précédemment. $r \geq 1$ et $t \geq 1$ sont des paramètres réels, $s$=0,1,... est un paramètre entier et $m$ et $n$ représentent les ordres des modes suivant les directions verticale et horizontale (respectivement).

Dans ce cas, l'inégalité conduisant à la fréquence fondamentale est la suivante :

$$\omega_o^2 \le \min_{r,s,t} \frac{\int_\Omega \mu\left((\frac{\partial \psi_2}{\partial x})^2 + (\frac{\partial \psi_2}{\partial z})^2\right)dxdz}{\int_\Omega \rho \psi_2^2(x,z)dxdz} \qquad (9)$$

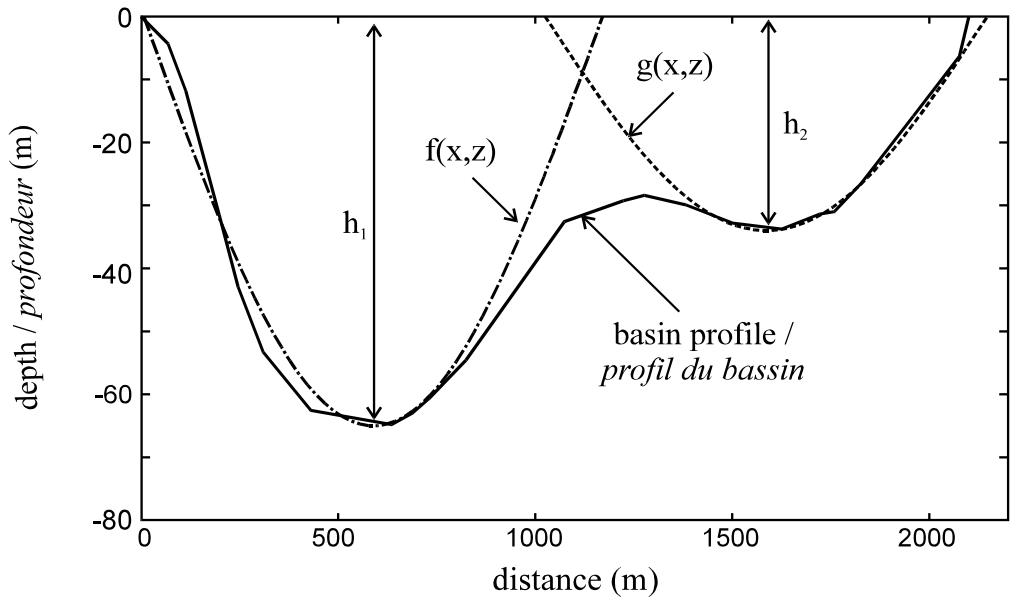

**Figure 1. Homogeneous geological profile for the simplified modal analysis.**
*Description du profil géologique homogène pour l'approche modale simplifiée.*

Les résultats sont obtenus pour le premier mode en considérant différentes valeurs du module de cisaillement : $\mu_1$=180 MPa (i.e. $V_S$=300 m/s), $\mu_2$=120 MPa (i.e. $V_S$=245 m/s) et $\mu_3$=90 MPa (i.e. $V_S$=212 m/s). Les résultats numériques obtenus par la méthode de Rayleigh sont comparés aux fréquences d'amplification maximale données par la méthode des éléments de frontière à travers une analyse explicite de la propagation des ondes (Semblat et al 00b).

TABLEAU I : *Comparaisons entre fréquences de référence obtenues par méthode propagative et fréquences fondamentales issues de l'approche modale simplifiée.*

| partie ouest | | partie est | |
|---|---|---|---|
| fréquence référence | fréquence fondamentale | fréquence référence | fréquence fondamentale |
| 1,35 | 1,50 | 2,42 | 2,86 |
| 1,30 | 1,23 | 2,13 | 2,34 |
| 1,13 | 1,07 | 1,75 | 2,02 |

## 3. Modal estimation of the fundamental frequency

### 3.1 Dependency of fundamental frequencies on shear modulus

D'après ces résultats, nous pouvons constater que les fréquences fondamentales obtenues par la méthode de Rayleigh sont en bon accord avec les fréquences de forte amplification déterminées par approche propagative. Les valeurs données dans le tableau I sont comparées pour la résonance de la partie profonde du bassin (ouest) et la partie peu profonde (est) et ce pour différentes valeurs du module de cisaillement. Pour un modèle de bassin homogène, la méthode de Rayleigh semble donc donner des résultats très intéressants qui sont non seulement en accord avec les valeurs obtenues à l'aide d'autres modèles mais également avec les résultats expérimentaux présentés en détail dans (Semblat et al 00b).

## 3.2 Comparison between simplified modal method and amplification curves

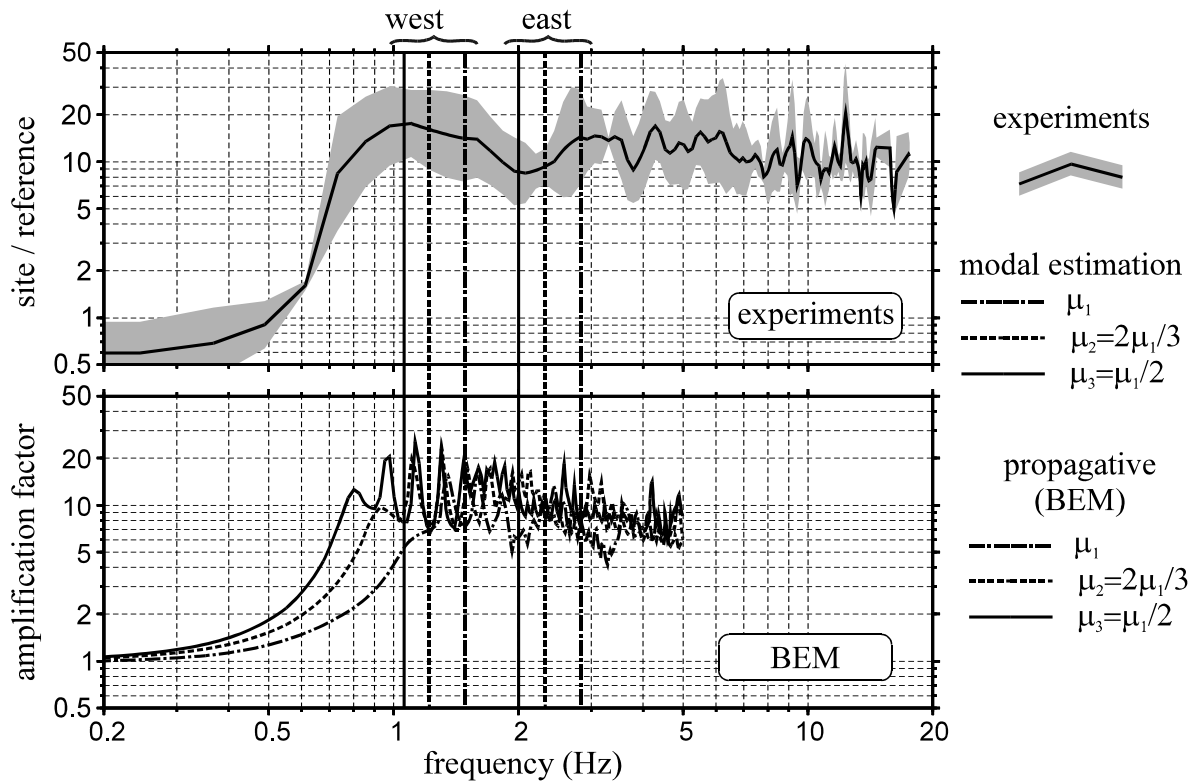

**Figure 2. Fundamental frequencies from the simplified modal approach (vertical lines) compared with experimental (top) and propagative BEM results (bottom).**
*Fréquences fondamentales estimée par méthode modale simplifiée et comparaison avec les mesures (haut) et les simulations par la méthode des équations intégrales de frontière (bas).*

## 4. Conclusion

L'analyse des effets de site par des approches modales est possible tant du point de vue qualitatif que quantitatif. Le site considéré dans cette note est un remplissage sédimentaire du centre de Nice pour lequel les mesures expérimentales prévoient des facteurs d'amplification variant entre 10 et 30 pour des fréquences comprises entre 1 et 2 Hz.

Les approches modales permettent de déterminer différentes fréquences propres d'un remplissage sédimentaire. Pour une application concrète de ces approches modales, quelques points devront encore être approfondis : la distribution de vitesse dans le remplissage, l'effet de l'amortissement...